# Development of a Vision System to Enhance the Reliability of the Pick-and-Place Robot for Autonomous Testing of Camera Module used in Smartphones

Hoang-Anh Phan
Research Center for Electronics and Telecommunications
*VNU University of Engineering and Technology*
Hanoi, Vietnam
anh.ph@vnu.edu.vn

Duy Nam Bui
Faculty of Electronics and Telecommunications
*VNU University of Engineering and Technology*
Hanoi, Vietnam
duynamk63uet@gmail.com

Tuan Nguyen Dinh
Faculty of Information Technology
*VNU University of Engineering and Technology*
Hanoi, Vietnam
ndinhtuan15@vnu.edu.vn

Bao-Anh Hoang
Research Center for Electronics and Telecommunications
*VNU University of Engineering and Technology*
Hanoi, Vietnam
anhhb@vnu.edu.vn

An Nguyen Ngoc
Faculty of Electronics and Telecommunications
*VNU University of Engineering and Technology*
Hanoi, Vietnam
ngocan@vnu.edu.vn

Dong Tran Huu Quoc
Faculty of Electronics and Telecommunications
*VNU University of Engineering and Technology*
Hanoi, Vietnam
dongtran.robotics@gmail.com

Ha Tran Thi Thuy
Faculty of Electronics Engineering
*Posts and Telecommunications Institute of Technology*
Hanoi, Vietnam
hatt@ptit.edu.vn

Tung Thanh Bui
Faculty of Electronics and Telecommunications
*VNU University of Engineering and Technology*
Hanoi, Vietnam
tungbt@vnu.edu.vn@gmail.com

Van Nguyen Thi Thanh
Faculty of Electronics and Telecommunications
*VNU University of Engineering and Technology*
Hanoi, Vietnam
vanntt@vnu.edu.vn

*Abstract*— **Pick-and-place robots are commonly used in modern industrial manufacturing. For complex devices/parts like camera modules used in smartphones, which contain optical parts, electrical components and interfacing connectors, the placement operation may not absolutely accurate, which may cause damage in the device under test during the mechanical movement to make good contact for electrical functions inspection. In this paper, we proposed an effective vision system including hardware and algorithm to enhance the reliability of the pick-and-place robot for autonomous testing memory of camera modules. With limited hardware based on camera and raspberry PI and using simplify image processing algorithm based on histogram information, the vision system can confirm the presence of the camera modules in feeding tray and the placement accuracy of the camera module in test socket. Through that, the system can work with more flexibility and avoid damaging the device under test. The system was experimentally quantified through testing approximately 2000 camera modules in a stable light condition. Experimental results demonstrate that the system achieves accuracy of more than 99.92%. With its simplicity and effectiveness, the proposed vision system can be considered as a useful solution for using in pick-and-place systems in industry.**

## I. INTRODUCTION

In modern manufacturing strategy, basic parts of a product are manufactured by separated vendors and then are assembled for the final product. With the continued development of autonomous systems, parts are mostly fabricated by autonomous systems to increase reliability, increase throughput and reduce fabrication cost. In this context, robots have been widely used for pick-and-place operation for autonomous assembly or testing purposes [1]–[5]. The pick-and-place operations normally are implemented in sequence by a program without an vison system for maximizing throughput [5], [6]. However, for testing of complex component like camera modules used in smartphones, which including optical components, electrical parts, and connectors, alignment them to the right place for making electrical connection and performance testing may not always success. The device under test can be damaged if having any error by mechanical interaction come from the incorrectly placement of the camera in the interfacing socket (Fig. 1). Therefore, development of a vision system to enhance the reliability of the pick-and-place robot arm for autonomous testing memory of camera module is in great demand. Vision systems are embedded to guide the robot in pick-and-place operation [7]–[9]. These system uses several approaches to implement this function. Markus et. al., has proposed a computer vision-controlled grid-placer robot that is able to automatically place grids during support film preparation [8]. The x-y arm design is based on a five-bar mechanism with the computer vision system based on a Hama Speak2 web camera and the control software is implemented in Python using the OpenCV Open-Source Computer Vision Library. The prototype operates with 90% success rate in preparing useful grids. Saurabh et. al., implemented a vision-based pick robot using a webcam and ABB IRB140 industrial robot [9]. A colour sorting algorithm developed in Matlab using image processing and computer vision toolbox. Object colour can be detected, and commands are provided to the controller to operate the robot successfully.

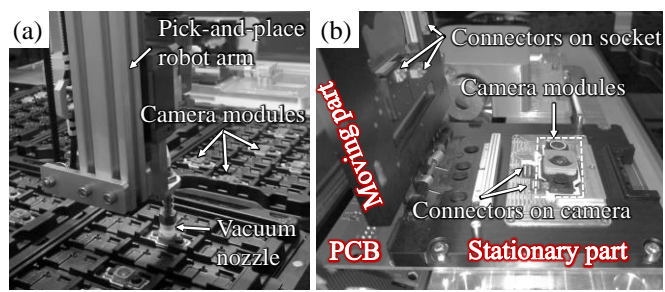

Fig. 1. System for testing functionality of the memory of camera module used in smartphones. (a) Pick-and-place robot arm for pick up device under test to testing position. (b) Interfacing socket with the camera module. Connector on camera module should be connected with the connector on the moving part of the socket during the test.

In the above-mentioned approaches, the vision system uses a quite complicated analysis process using computer vision library. In our work, we propose an effective vision system for determination of the presence and accuracy of camera module placement. The system includes the basic hardware and image processing algorithms. The hardware system is designed optimizing and modularization while the image processing algorithms focus on analysing the characteristics of camera module image to find the value data sets, from that determining the presence and correct placement of the camera components on the trays. The results are the machine with less processing time and mistake avoidance that can damage the camera module during real time.

## II. Autonomous System for Testing Memory of Camera Module

The hardware structure of the main system is designed optimizing and modularization. Fig. 2 shows the block diagram of the hardware structure, consisting of four main parts: (I) a console and display unit, (II) a pick-and-place robot arm with vacuum tool department, (III) a memory testing department and (IV) vision system department to enhance the reliability of the Pick-and-Place robot arm.

The pick-and-place mechanical arm constructed from a cartesian coordinate robot arm which is driven by 3 servo motors by the MCU 1 (Arduino Mega). A pneumatic pump which attached to the robot arm is used for picking and placing the camera modules. Memory of camera module is tested by writing and reading operation through the interfaccing socket and testing circuit with standard protocol. The test result signal will be transferred to MCU 2 as the quality of the camera module data. The Vision inspection department includes two separate cameras (Raspberry Pi Camera Modules V2 8 MP) to detect the presence and the placement accuracy of the camera modules. Each inspection camera is controlled by an embedded computer (Raspberry Pi 4 model B) and they send results to MCU 2 and MCU 3 to decide the next motion. MCU 1 is the center controller for receiving and processing the data from the two inspection cameras.

In this work, vacuum pressure and precision nozzles are used to enable camera module under test (CMUT) placement. There are several distinct stages to the autonomous CMUT memory testing process as described as follows. First, a selected CMUTs are withdrawn from the feeding tray by a vacuum nozzle. The CMUT is transferred from the picking location to the alignment zig position. The CMUT is lowered and released by the nozzle. The alignment zig with trapezoid shape can align the CMUT to its right position. The robot arm picks the CMUT again and brings it to the socket position with pre-program coordinator. The socket the is closed to make good contacts between the CMUT and testing department for testing the functionality of the memory. After the test finishing, the CMUT is picked and bring back to its position in the feeding tray. The process is restarted for another module and the process is repeated for every camera module in the feeding tray. In case the CMUT does not pass the test, it will be placed in specific position in a tray for rework.

Although the CMUTs are aligned by the alignment zig, it still occurs the inaccuracy placing of the CMUT in the socket in some cases due to the complexity of the structure of the camera module. In that case, when the socket is closed to make good contact with the memory inspection department, the socket can damage the connector or even the CMUT. A vision system can be embedded to monitor position of the CMUT before closing the socket will increase the reliability of the system.

## III. Method to Determine the Presence of a Camera Module in The Feeding Tray and the Placement Accuracy of yhe Camera Module in The Memory Testing Socket

The presence of a CMUT in the feeding tray and placement accuracy of a camera module in the memory testing socket can be determined through property of the object in the image. First, the image of the CMUTs in the feeding tray are sampled and their histogram are calculated. Depending on its histogram, the present of a CMUT in the feeding tray or placement accuracy of the CMUT in the memory testing socket can be determined. Details will be described in the following sections.

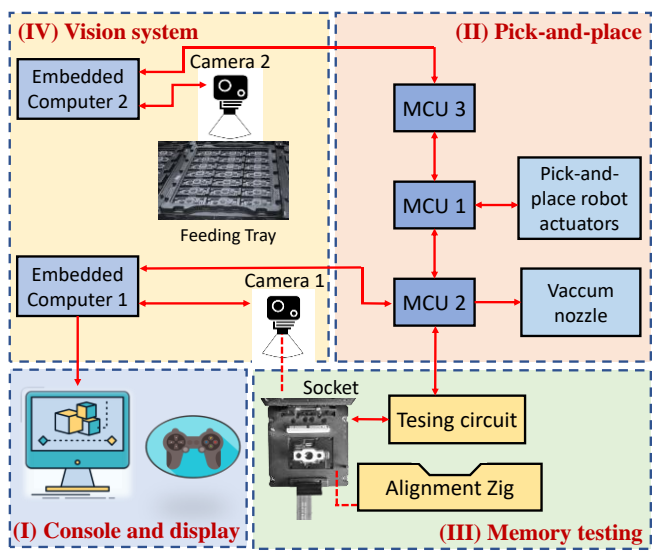

Fig. 2. Block diagram of the pick-and-place robot arm based system for autonomous testing memory of camera module.

## A. Determination the Presence of Camera Modules in Loading Trays

Fig. 3(a) shows an image of feeding trays in working space. The zoom-in images show clearly that there are more black pixels in an empty image (Fig. 3(b)) than with camera module image (Fig. 3(c)).

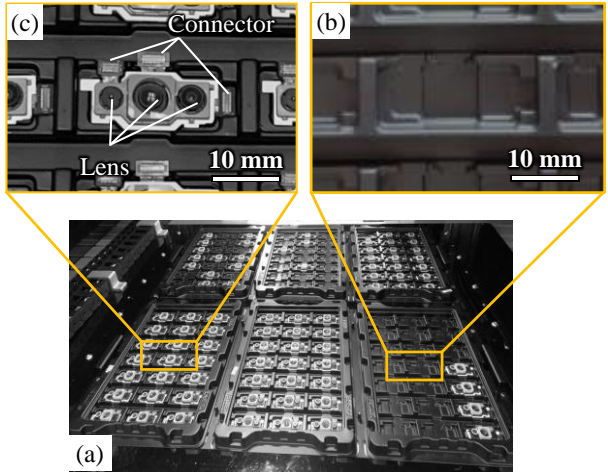

Fig. 3. Working space with feeding trays containing camera modules used in smartphones. (a) Total working space view. Zoom-in at the positions without (b) and with the camera modules (c).

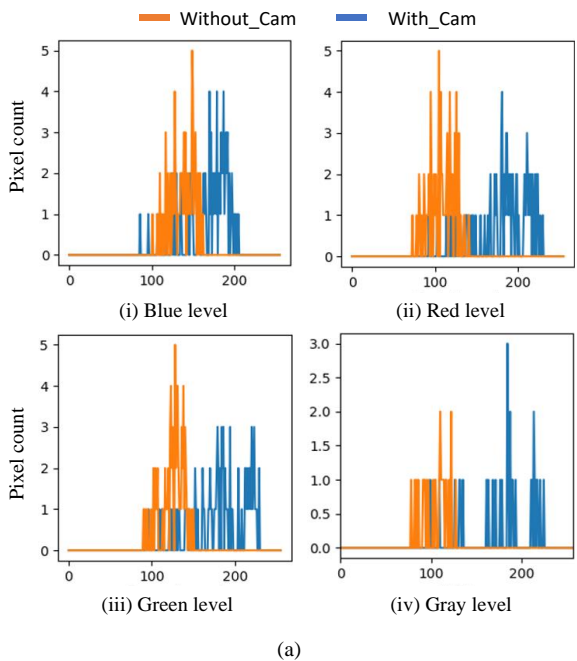

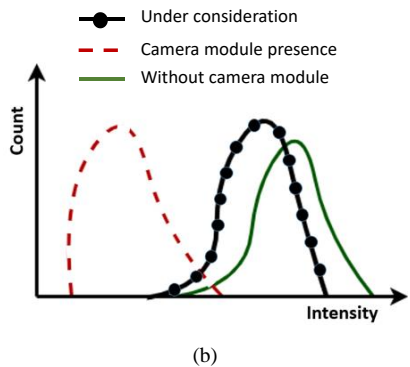

Fig. 4. The distribution area of image: (a) The histogram of "w/i cam" and "w/o cam" image, (b) The graph of distribution area.

The properties of these two images are identified in their histogram in Fig. 4. In order to show the image properties clearly, the histogram is shown by the colour image with three channels (Red, Green and Blue) and by the Gray image (Fig. 4(a)). The name With_Cam and Without_Cam are used for the images with camera module in the tray and without camera module in the tray, respectively. The pixel value distribution area of With_Cam is concentrated at a value higher than the area of Without_Cam for all results. Fig. 4(b) shows the distribution area of each image in the graph form. This graph is the set of values obtained after sampling the image.

From the above properties of with and without camera module images, the block diagram of Camera module presence approach is proposed as Fig. 5. The detection process is divided into two phases:

- Reference data sampling: information of With_Cam and Without_Cam in the tray is collected and these two data sets are stored to use as reference for determining the presence of camera module in the tray.

- Sampling, comparison and determination: during testing the CMUT, the vision system will capture the tray on the working space. The image is cropped and calculated its value. Each is then compared to the reference value obtained in phase 1. The result of the determination has a sequence of 1/0 corresponding to Yes/No in the order of modules on the trays.

The algorithms of these two mentioned phases are shown in Table I and II as follows.

TABLE I. ALGORITHM FOR SAMPLING AND GETTING SAMPLE VALUE SETS FOR DETERMINING THE PRESENT OF CAMERA MODULE

| Algorithm 1: Sampling and getting sample value sets |
| --- |
| **Result**: Sample value set<br>Sampling the images;<br>Cutting sub-image to with_cam_dataset and without_cam_dataset;<br>**for** *i in len(with_cam_dataset)* **do**<br>    hist_with_cam[i]=histogram(with_cam_dataset[i]);<br>    value_with_cam[i]=mean(hist_with_cam[i]);<br>**end**<br>**for** *i in len(without_cam_dataset)* **do**<br>    hist_ without_cam[i]=histogram(without_cam_dataset[i]);<br>    value_ without_cam[i]=mean(hist_without_cam[i];<br>**end** |

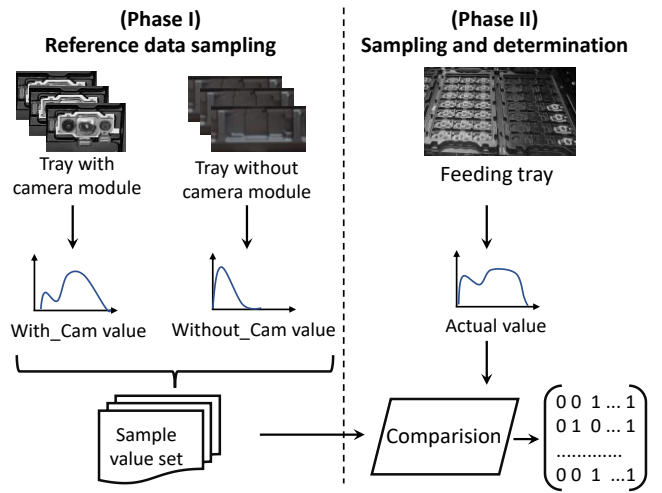

Fig. 5. Block diagram of the approach to detech of the presence of camera modules in the feeding tray.

TABLE II. ALGORITHM FOR DETERMINING THE PRESENCE OF CAMERA MODULE

**Algorithm 2: Determination the presence of camera module**
**Result**: True/False
Capture the images;
Cutting sub-image to unknown_dataset;
**for** *i in len(unknown_dataset)* **do**
    hist_unknown[i] = histogram(unknown_dataset[i]);
    value_unknown[i] = mean(hist_unknown[i]);
    a = value_unknown[i] – value_with_cam[i];
    b = value_unknown[i] – value_without_cam[i];
    **if** */a < b/* **then**
        True;
    **else**
        False;
    **end**
**end**

*B. Determination of the Aaccuracy of Camera Module Placement*

Differ from the case detecting the presence of camera module in the feeding tray, the camera module surely placed in the socket. However, it may have misalignment for some reason and not fitting with the designated shape on the socket. To determine the accuracy of camera module placement in testing socket, the With_Cam image is sampled 150 times to find the expecting value. Fig. 6(a) shows the distribution of obtained values after sampling while the distribution graph is presented in Fig. 6(b) with normal distribution. Therefore, the mean and standard deviation values are calculated. The threshold for the distribution is set at a confidence level of 95%. The value obtained within this range will be considered to be correct placement. The mean $\bar{x}$, variance $s^2$ and confidence interval (SI) are calculated by the equations (1) to (3) [10], [11].

$$\bar{x} = \frac{1}{n}\left(\sum_{i=1}^{n} x_i\right) = \frac{x_1 + x_2 + x_3 + \cdots + x_n}{n} \quad (1)$$

$$s^2 = \frac{\sum_{i=1}^{n}(x_i - \bar{x})^2}{n-1} \quad (2)$$

$$CI = \bar{x} + z\frac{s}{\sqrt{n}} \quad (3)$$

where, *n* is the sample size; $x_1, x_2, ... x_n$ are the measures values; *s* is the standard deviation; and *z* is the confidence level value.

Fig. 7 describes the block diagram of the placement accuracy detection process. It is also divided into two phases:

• Data sampling: the number of images used for sampling is 30. This value is large enough to calculate the expecting values with high confidence. From the obtained images, the part of the image which contains modules in the socket will be cropped to calculate the mean and standard deviation to store as the value sets.

• Determining: after being placed in the interfacing socket, the vision system will capture the image of the socket. Then image of the camera module will be cropped and compared to the reference mean and standard deviation values taken in data sampling phase. Determination result will be in the binary form, corresponding to Correct/Incorrect in placing of the CMUT in the socket.

The algorithms to determine the accuracy of camera module placement are shown in Algorithm 3 and Algorithm 4 as below.

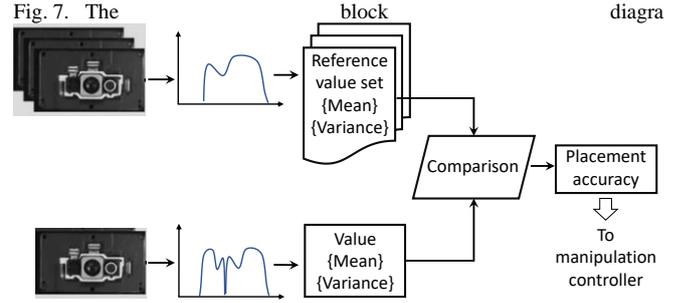

Fig. 7. The block diagram of the placement accuracy detection process

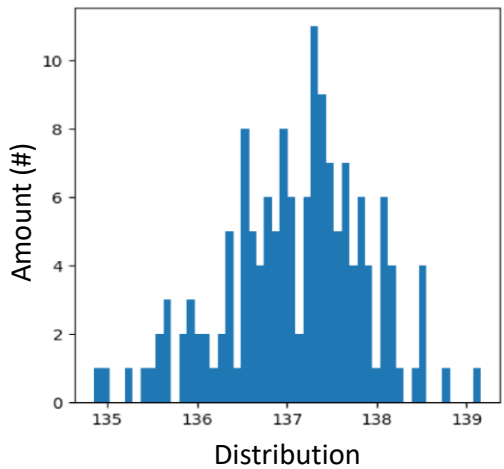

(a)

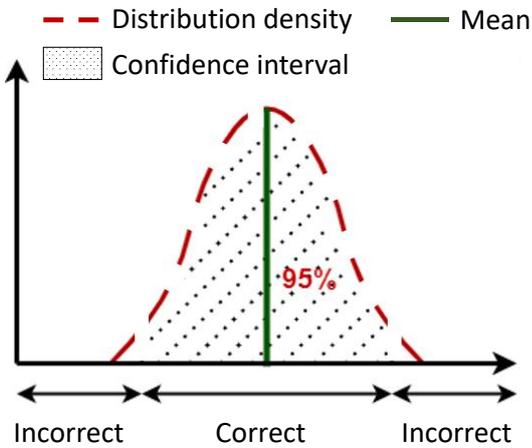

(b)

Fig. 6. The distribution area of image: (a) The distribution of the obtained valued after sampling, (b) The graph of distribution area

TABLE III. ALGORITHM FOR SAMPLING AND GETTING SAMPLE VALUE SETS FOR DETERMINING THE PLACEMENT ACCURACY

**Algorithm 3: Sampling and getting sample value sets**
**Result**: Sampling value set [mean,variance]
Sampling the image (more than 30 times) and get dataset;
**for** i in len(dataset) **do**
    hist[i] = histogram(dataset[i]);
    value[i] = mean(hist[i]);
**end**
mean_value = sum(value[i]/len(dataset);

TABLE IV. ALGORITHM FOR DETERMINING OF THE ACCURACY OF CAMERA COMPONENT PLACEMENT

| Algorithm 4: Determination of the accuracy of camera component |
|---|
| **Result**: True/False or Correct/Incorrect)] |
| Capture the images; |
| CI = 1.96; |
| hist = histogram(image); |
| value = mean(hist); |
| **if** abs(value – mean_value) ≤ CI*var_value **then** |
|     True; |
| **else** |
|     False; |
| end |

var_value = sqrt(sum((value – mean_value)2)/(n-1))

## IV. EXPERIMENT RESULTS AND DISCUSSION

The experiments were conducted 30 times, with different number of camera modules, divided into 4 trays in each experiment. The total number of camera modules used in the experiment is approximately 2000. Sensitivity and specificity of detection of the presence of camera module in the feeding tray and the accuracy placement of camera module in the test socket of the total 2000 camera modules were calculated to evaluate the performance of the vision system.

With 84 trays with camera modules inside were feed to the system. The order of the determination of the presence of the camera module in feeding tray will be performed from left to right, top to bottom. A program was developed to observe and display the position of missing camera modules on the tray. Fig. 8 shows the results of different experiment cases with the present and absence of the camera module on the tray. On the display screen, difference colours represent the corresponding position with the camera modules presence and absence of the camera modules. Obtained result show that the system can correctly identify the presence of camera module in the feeding trays.

The Actual – Predict model is used to evaluate the effectiveness of the proposed method. The True Positive (TP), False Positive (FP), True Negative (TN) and False Negative (FN) based on the un-normalized confusion matrix as in Table V [12].

TABLE V. STATISTICS OF THE ACTUAL – PREDICT MODEL

|  |  | Predict | |
|---|---|---|---|
|  |  | *Positive* | *Negative* |
| **Actual** | *Positive* | TP | FN |
|  | *Negative* | FP | TN |

From the Actual – Predict model, the accuracy, precision and recall can be calculated as bellows [13-14].

$$Accuracy = \frac{TP + TN}{TP + TN + FP + FN} \quad (4)$$

$$Precision = \frac{TP}{TP + FP} \quad (5)$$

$$Recall = \frac{TP}{TP + FN} \quad (6)$$

Experimental results are shown in Table V. The incorrect placements are set on purpose. A large number of With_Cam images are used in this case. From the results in Table VI, some criteria are found following the equations (4-6) and shown as in Table VII. The results show the effectiveness of the proposed method with high value in all criteria.

TABLE VI. STATISTICS OF THE RESULTS OBTAINED FROM EXPERIMENTS

|  |  | Predict | |
|---|---|---|---|
|  |  | *True* | *False* |
| **Actual** | *True* | 4334 | 2 |
|  | *False* | 13 | 13641 |

TABLE VII. THE ASSESSMENT CRITERIA

| Parameter | Criteria |
|---|---|
| Accuracy | 0.9992 |
| Precision | 0.9970 |
| Recall | 0.9995 |

Experimental results show that the proposed vision system has significantly improved the reliability of the memory of camera module autonomous testing system. The vision system is quite simple, utilizing of basic hardware and effective algorithm with short processing time. The vision system does not affect to the throughput of the machine due to its insignificant calculating time to compare with times consuming to bring the camera module from feeding tray to socket position and testing time of the memory of the camera module. The presence of this vision system helps to enhance the reliability of the machine and it would help to avoid damaging the device under test when having error in the placement process.

## V. CONCLUSIONS

The paper proposed and implemented the vision system including hardware and algorithm to detect the presence of the camera module in feeding tray and the placement accuracy of the camera module in interfacing socket to enhance the reliability of the pick-and-place system for autonomous testing memory of camera module. The effectiveness of the proposed system is verified by experimental results. The proposed system has improved the performance of the machine with less affect to the throughput of the machine while increase its reliability. The proposed approach can be extended to other pick-and-place machine in autonomous testing systems.

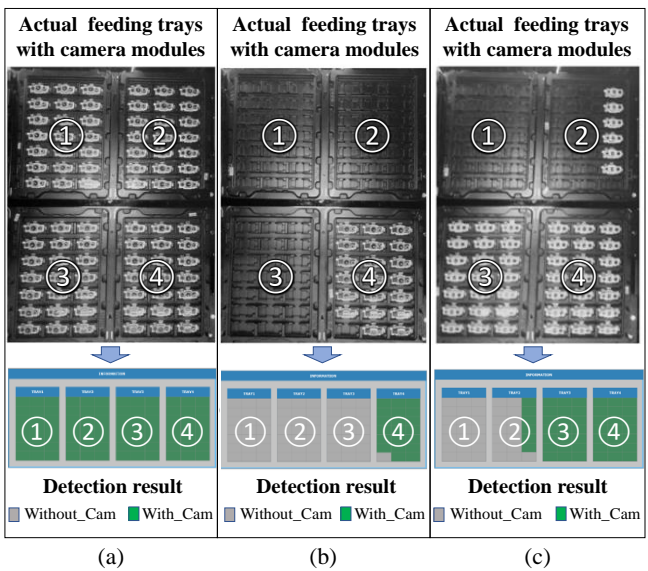

Fig. 8. The result of the determination of the presence of the camera module.

ACKNOWLEDGMENTS

This work has been supported by VNU University of Engineering and Technology under project number CN20.31.